\documentstyle[12pt,epsfig]{article}

\newcommand{\lsim}{\stackrel{<}{\sim}}

\newcommand\GeV{\,\mbox{GeV}}

\newcommand\alsq{\alpha_s(Q^2)}
\newcommand\alsn{\alpha_s(Q^2_0)}
\newcommand\alsm{\alpha_s(M^2_Z)}
\newcommand\alst{\alpha_s(t)}
\newcommand\gli{\widetilde{g}}
\newcommand\gl{\tilde{g}}

\newcommand\Pvec{\, \mbox{\boldmath $P$}}
\newcommand\Rvec{\, \mbox{\boldmath $R$}}

\begin{document}
\thispagestyle{empty}

\mbox{}
\begin{flushright}
DESY 94-008\\
January   1994\\
\end{flushright}
\vspace*{\fill}
\begin{center}
{\LARGE\bf Do Deep Inelastic Scattering Data Favor}

\vspace{2mm}
{\LARGE\bf a Light Gluino?}
\\

\vspace*{20mm}

\large
\begin{tabular}[t]{c}
Johannes Bl\"umlein~~and~~James Botts
\vspace*{10mm}
\\

{\it DESY--Institut f\"ur Hochenergiephysik,}\\
{\it Platanenallee 6, D--15735 Zeuthen, FRG}\vspace*{10mm} \\
\end{tabular}
\end{center}
\vspace*{\fill}
\begin{abstract}
\noindent
A next--to--leading order QCD analysis of deep inelastic scattering data
is performed allowing for contributions due to a light gluino.
We obtain the values of $\alpha_s(M_Z^2) \pm
\delta \alpha_s^{stat}
= 0.108 \pm 0.002,
0.124 \pm 0.001, 0.145 \pm 0.009$ for QCD, SUSY QCD with a
Majorana gluino and  a Dirac gluino respectively.
The value of $\alsm$
obtained in SUSY QCD with a Majorana gluino
best agrees with the direct measurements of $\alpha_s(M_Z^2)$ at
LEP.
\end{abstract}
\vspace*{\fill}
\newpage
\noindent
The most precise measurements of $\alpha_s(Q^2)$ are provided by
deep inelastic lepton--hadron scattering  and $e^+e^-$
experiments and extend over many generations of scale providing
a test of QCD~\cite{REV}. The results of the QCD analyses performed in
the different deep inelastic scattering experiments may be expressed
in terms of a value for $\alpha_s(M_Z^2)$
(by a next--to--leading order (NLO) relation, for example)
and thus directly compared with the results obtained
in $e^+e^-$ experiments at LEP.  Different values
of $\alsm$ are obtained from deep inelastic and $e^+e^-$ experiments.
The question has arisen
whether this difference might be
due to the
presence of a light gluino~\cite{ELLIS,KJ} for which windows
remain open in the range
$m_{\gli} \lsim 5 \GeV$~\cite{PDG,CLAV}\footnote{Different
possibilities for a
dedicated search for light gluinos have been also
proposed in~\cite{ELLIS,PROPO,RS} recently.
Light gluinos are also favored in some
approaches in string theory (see~\cite{LUEST}).}.

Unlike  direct measurements of $\alsm$ in LEP experiments, the QCD analysis
of deep inelastic scattering data requires an assumption on the
$\beta$--function to solve the evolution equations.
As shown in~\cite{KJ}
the re--evaluation of $\alsm$ resulting from the DIS--data under the
assumption of the NLO $\beta$--function in the presence of
a light Majorana gluino~\cite{BESUSY} yields a better agreement with data.

However, the effect of a light gluino in the mass range
$m_{\gli} \lsim 5 \GeV$
on the scaling violations of
structure functions is {\it not only} due to a modification of the
$\beta$--function, but also due to modified evolution equations
containing even different parton densities.
Since the allowed gluino mass is rather small, a thorough study of this
effect requires a complete
NLO QCD analysis of the deep
inelastic data including light gluinos\footnote{The effect of a
gluino with $m_{\gli} \sim 5 \GeV$ on data expected in the {\it high}
$Q^2$
range at HERA, decoupling at low $Q^2$ range, was
studied in~\cite{RS} in LO recently.}.
This is the purpose of the present paper.

\vspace*{1cm}
\noindent
\large {\bf
The Running Coupling Constant}
\normalsize

\vspace*{2mm}
\noindent
Before we study the evolution of deep inelastic structure functions
we summarize the effect of a gluino on $\alsq$. The renormalization
group equation for $\alpha_s$
\begin{equation}
\frac{\partial \alpha_s}{\partial \log \mu^2} =
- \frac{\beta_0}{4 \pi}  \alpha_s^2
- \frac{\beta_1}{(4 \pi)^2} \alpha_s^3
- \frac{\beta_2}{(4 \pi)^3} \alpha_s^4
+ ...
\end{equation}
yields the solution
\begin{equation}
\label{eq1}
\frac{1}{\alsq} = \frac{1}{\alsn} + \frac{\beta_0}{4 \pi} \log \left
( \frac{Q^2}{Q_0^2} \right )
+ \Phi^{(n)}(\alsq; \beta_i)
- \Phi^{(n)}(\alsn; \beta_i)
\end{equation}
with the function\footnote{Because
the explicit expression for (\ref{eqPHI})
depends on the size of $\beta_{0,1,2,...}$ $\Phi^{(n)}(x; \beta_i)$
is understood to be a complex valued function.}
$\Phi^{(n)}(x; \beta_i)$  having the form
in next--to--next to leading order (NNLO)
\begin{eqnarray}
\label{eqPHI}
\Phi^{(n)}(x; \beta_i) &=& - \frac{\beta_1}{8 \pi \beta_0} \ln \left |
\frac{16 \pi^2 x^2}{16 \pi^2 \beta_0 + 4 \pi \beta_1 x + \beta_2 x^2}
\right | \nonumber \\
&+& \frac{\beta_1^2 - 2 \beta_0 \beta_2}{8 \pi \beta_0 \sqrt{4 \beta_2
\beta_0 - \beta_1^2}} \arctan \left( \frac{2 \pi \beta_1 +\beta_2 x}{
2 \pi \sqrt{4 \beta_0 \beta_2 - \beta_1^2}} \right )
+ {\cal O}(\beta_3)
\end{eqnarray}
In the limit $\beta_2 \rightarrow 0$ (NLO)
one obtains
\begin{equation}
\Phi^{(1)}(x; \beta_i) = - \frac{\beta_1}{4 \pi \beta_0} \ln \left [
\frac{\beta_0^2  x}{4 \pi \beta_0 +  \beta_1 x}
\right ] + C
\end{equation}
where we choose $C = 0$ following~\cite{LAM1}\footnote{
This choice is often
called  definition of
$\alsq$ in the
$\overline{\rm MS}$~scheme, although $\beta_{0,1}$ are  scheme
independent and only $\beta_2$ eq.~(\ref{BEEQ}) refers to a
$\overline{\rm MS}$ result. Other possible choices are discussed
in~\cite{LAM2}.}.
The coefficients $\beta_i$ are given by~\cite{BE,BESUSY}
\begin{eqnarray}
\label{BEEQ}
\beta_0 &=&  11 - \frac{2}{3} N_f - 2 N_{\gli} \nonumber
\\
\beta_1 &=&  102 - \frac{38}{3} N_f - 48 N_{\gli}
\nonumber \\
\beta_2 &=& \frac{2857}{2} - \frac{5033}{18} N_f + \frac{325}{54} N_f^2
+ \widetilde{\beta}_{2,\gli}(N_f, N_{\gli}) \ .
\end{eqnarray}
$ \widetilde{\beta}_{2,\gli}(N_f, N_{\gli})$
describes the gluino
contribution to $\beta_2$.
$N_f$ denotes the number of active quark flavours and
$N_{\gli} = 0,1,2$ refers to QCD, supersymmetric QCD with
Majorana gluinos, and Dirac gluinos, respectively.
Eq.~(\ref{eq1}) can be solved iteratively. If $\alsn$ is chosen by
$\alpha_s(M^2_Z)$ one can calculate $\alsq$ as a function of $\beta_0$
and $\beta_1$, and the dependence on $\Lambda$ is
implicit only through $\alpha_s(Q^2_0)$.
The $\Lambda$ dependence
can be made explicit through the definition
\begin{equation}
\label{eqLAMZ}
\Lambda   :=  Q_0 \exp \left \{
- \frac{2 \pi}{\beta_0 \alsn} + \frac{\beta_1}
{2 \beta_0^2}
 \log \left [
\frac{\beta_0^2 \alsn}{4 \pi \beta_0 + \beta_1 \alsn} \right ] \right \}
\end{equation}
in the $\overline{\rm MS}$ scheme in NLO\footnote{A
corresponding relation in NNLO
follows from (\ref{eq1},\ref{eqPHI}) directly.}.
Eq. (\ref{eqLAMZ})  relates the values of $\Lambda$ and $\alpha_s$
at a given scale {\it directly}.

When quark or gluino
mass--thresholds  $\mu \approx 2 m_i$
are passed, $\alsq$ is kept continuous
as the values of $\beta_i$ change.
To be more precise, running mass effects
should be accounted for when passing quark or gluino mass
thresholds~\cite{THR}. We will neglect these effects in the present
analysis as small\footnote{
 As has been shown
recently \cite{SHIRK} this
modification leads to a shift $\Delta \alsm \sim 0.001$ unless a quark
mass threshold coincides with that of the gluino.}.

In figure~1 a comparison of the NLO
solutions (\ref{eq1})
using $\alsn = \alsm = 0.122$
for $N_{\gli} = 0,1,2$, $m_{\gli} = 0, 3, 5 \GeV$,
 and different measurements   of $\alpha_s$ is shown. Here we assumed
$m_c = 1.6 \GeV$ and $m_b = 4.75 \GeV$  for the values of the heavy
quark masses. For the case of QCD also the curve obtained in NNLO is
given.
That a light Majorana gluino is an apparently better description
of $\alpha_s(Q^2 > 16 \ GeV^2)$ than QCD has been previously observed
in \cite{KJ}.
Calculating
confidence levels for the depicted experimental data in the complete
$Q^2$ range\footnote{The $\alpha_s$ value determined from $c \overline{c}$
$1P-1S$
mass splitting using a lattice calculation~\cite{LAT} was excluded.}
from the $\chi^2$
values for the three
different hypothesis ${N_{\gl} = 0,1,2}$, however,
one obtains  78 \% CL for QCD, 50--83 \% CL for  the case of a
Majorana gluinos ($0 < m_{\gl} < 5 \GeV$), and
$<$ 0.5 \% CL
 for the case of a Dirac
gluino. This estimate, however, assumes, that the $\alpha_s$ values
have been extracted in a way which is essentially insensitive to
the value of $N_{\gl}$, since most of these values were obtained
using QCD matrix elements as an input.

\vspace*{1cm}
\noindent
\large {\bf
QCD Analysis}
\normalsize

\vspace*{2mm}
\noindent
Scaling violations of the structure functions  measured in different
deep inelastic scattering experiments can be described\footnote{The
data analysis performed on this basis will include the data currently
measured at HERA~\cite{EXP1} also. Although these are
small $x$ data they also exhibit logarithmic scaling violations.}
 by the following
set of NLO evolution equations including the effect of a light gluino,
\begin{eqnarray}
\label{eqevo}
\frac{d}{dt} q_i^{(-)}(t,x) &=& \left [ P_{NS}^{(0)}(x) + \frac{\alst}
{2 \pi} \widetilde{P}_{NS,-}^{(1)}(x) \right ] \otimes
q_i^{(-)}(t,x) \nonumber\\
\frac{d}{dt} \widetilde{q}_i^{(+)}(t,x)
&=& \left [ P_{NS}^{(0)}(x) + \frac{\alst}
{2 \pi} \widetilde{P}_{NS,+}^{(1)}(x) \right ] \otimes
\widetilde{q}_i^{(+)}(t,x)
\nonumber \\
\frac{d}{dt} \left [
\begin{array}{c}q^{(+)}(t,x) \\
                     G(t,x)     \\
                  \gli(t,x)
\end{array}
\right ] &=& \left [ \Pvec^{(0)}(x) + \frac{\alst}{2 \pi} \Rvec(x)
\right ] \otimes \left [
\begin{array}{c}q^{(+)}(t,x) \\
                     G(t,x)    \\
                  \gli(t,x)
\end{array}
\right ] \ .
\end{eqnarray}
The gluino density $\gli(t,x)$ emerges in the singlet evolution equation.
In the description of the singlet and non--singlet distributions we
assume that the squark contributions decouple~\cite{DECOUP} due to
the current experimental limit $m_{\tilde{q}} > 74 \GeV$~\cite{PDG}.
The functions
\begin{eqnarray}
\label{eqRP}
R_{\pm}(x) &=& P_{NS, \pm}^{(1)}(x)
- \frac{\beta_1}{2 \beta_0} P_{NS}^{(0)}(x) \\
\Rvec(x) &=& \Pvec^{(1)}(x)
- \frac{\beta_1}{2 \beta_0} \Pvec^{(0)}(x)
\end{eqnarray}
are expressed by the splitting functions $P^{(k)}(x)$ up to NLO.
Note, that the leading order  matrix elements $P^{(0)}_{q\gli}(x)$ and
$P^{(0)}_{\gli q}(x)$ of $\Pvec^{(0)}(x)$ vanish.

The different combinations of quark densities
in (\ref{eqevo}) are
\begin{eqnarray}
q_i^{(\pm)}(t,x) &=& q_i(t,x) \pm \overline{q}_i(t,x) \nonumber \\
q^{(+)}(t,x) &=&  \sum_{i=1}^{N_f} q^{(+)}_i(t,x) \nonumber \\
\widetilde{q}^{(+)}_i(t,x) &=&
q^{(+)}_i(t,x) -  \frac{q^{(+)}(t,x)}{N_f}  \ .
\end{eqnarray}
The evolution variable is
$t = -(2/\beta_0) \log[\alsq/\alsn]$, with
$Q_0$ defining  the scale at which the input distributions
$q_i^{(\pm)}(x,Q^2_0), G(x,Q^2_0)$, and $\gli(x,Q^2_0)$  are parametrized.
$\otimes$ describes the Mellin convolution.

The splitting functions  $P^{(k)}_{ij}(x)$ in (\ref{eqevo},\ref{eqRP})
were given in \cite{CE,PSUSY} where in NLO the
$\overline{\rm DR}$~scheme was used. In the present analysis
these results were translated into the $\overline{\rm MS}$~scheme.

The QCD analysis was performed with a modified version of
the CTEQ parton distribution evolution
program~\cite{CTEQP}. The input shapes for the parton distributions
at $Q^2_0$ were chosen to be
\begin{eqnarray}
\label{eqpar}
xu_v(x) & = & A_u x^{\alpha_u} (1  - x)^{\beta_u}(1 + \gamma_u x)
\nonumber \\
xd_v(x) & = & A_d x^{\alpha_d} (1  - x)^{\beta_d}(1 + \gamma_d x)
\nonumber \\
x(\overline{d} + \overline{u})(x) & = &
A_+ x^{\alpha_+} (1  - x)^{\beta_+}(1 + \gamma_+ x)
\nonumber \\
x(\overline{d} - \overline{u})(x) & = &
  - x^{\alpha_-} (1  - x)^{\beta_-}(1 + \delta_- \sqrt{x} +
\gamma_- x)
\nonumber \\
xs(x) &=& A_s x^{\widetilde{\alpha_s}} (1 - x)^{\beta_s}
\nonumber \\
xG(x) &=& A_G x^{\alpha_G} (1 - x)^{\beta_G}(1 + \gamma_G x)
\nonumber \\
x\gli(x) &=& A_{\gl} x^{\alpha_{\gl}} (1 - x)^{\beta_{\gl}}
\end{eqnarray}
The sensitivity of the parameters in ({\ref{eqpar}}) to the fit
results varies, and some of them may be fixed or can
be related to one another.
The 10--parameter
parametrization used in our analysis was the result of a study to minimize
the number of shape parameters needed to fit the current
set of experimental inclusive data in a global QCD analysis.
This is desirable as fewer shape parameters are clearer to
analyze and reduce spurious correlations between themselves.
We set $\alpha_u = \alpha_d = 0.5$,
$\beta_u = 4.0$, $\beta_d = 3.0$, $\beta_+ = \beta_- \equiv \beta_G + 1$,
$\alpha_+ \equiv \alpha_G$, and $\gamma_+ \equiv \gamma_G$.
The sum rules
$\int_0^1 u_v(x) dx = 2$, $\int_0^1 xd_v(x) dx = 1$, and momentum
conservation reduce by three the number of input fitting parameters.
The parameters of the strange quark distribution have been determined
from the CCFR dimuon data~\cite{DIMU} as $A_s = 0.114$,
$\tilde{\alpha}_s = -0.114$, and $\beta_s = 6.87$ and are treated as fixed.
Only for $m_{\gli} \le Q_0 \equiv 1.6 $ GeV do we have
an initial three parameter nonzero gluino distribution.
Otherwise
the gluino distribution was radiatively generated as were the heavy flavor
distributions.

We allowed the relative normalizations of the different experimental
data sets (seven parameters) to float under the fit
subject to suitable constraints reflecting experimental normalization
uncertainty in the mean.

$\Lambda$ and the parameters of eq.~(\ref{eqpar})
were determined by a MINUIT~\cite{MINUIT} minimization
of a $\chi^2$--distribution
using the following data sets: SLAC $F_2^{ep}, F_2^{ed},
F_2^{ed}/F_2^{ep}$, NMC $F_2^{\mu d}, F_2^{\mu p},
F^{\mu n}_2/ F_2^{\mu p}$ for $E_{\mu} = 90~{\rm and}~280 \GeV$,
BCDMS $F_2^{\mu d}, F_2^{\mu p}$, CCFR $F_2^{\nu}, xF_3^{\nu}$, and
ZEUS and H1 $F_2^{ep}$ \cite{EXP,EXP1}
as input. The statistical and systematical experimental errors were
added in quadrature, and relative systematic effects between the
different experiments were accounted for by normalization factors
determined in the fit.
For a given gluino mass hypothesis it was demanded, that
$W^2 = Q^2(1 -x)/x > 4 m_{\gl}^2$ leading to a constraint
on the data used.
For the comparison of different possible hypotheses we considered
the cases $N_{\gli} = 0$ (QCD), and $N_{\gli} = 1,2$ with
$m_{\gli} = 0, 3,~{\rm and}~5 \GeV$. Table~1 summarizes the results
on $\Lambda$ and $\alsm$ and table~2 contains the shape parameters
determined for some characteristic cases
and the fitted
relative normalizations of the different data sets used in the
analysis. The latter parameters values differ by only up to 5 \%
from unity.
Note, that the values
of $\Lambda_{\overline{\rm MS}}^{(5)}$ (the value above the $b$--quark
threshold) and $\alsm$ given in table~1
are directly related by eq.~(\ref{eqLAMZ}) and {\it no} reference to a
"typical" value of $\langle Q^2 \rangle$ characterizing the analysed
data sets is needed\footnote{Such an assumption would be
required for an
illustration of the obtained result in the way shown in figure~1, but
will not be given due to the uncertainty in the abscissa.}.

\vspace*{1cm}
\noindent
\large {\bf
Discussion}
\normalsize

\vspace*{2mm}
\noindent
The values of $\chi^2 / {\rm NDF}$ obtained in the analysis of
all the deep inelastic scattering data
do not differ significantly (see table~1).
Thus, the analysis does not yield
an a priori preference to one of the cases $N_{\gli} = 0,1,2$.
The fitted values of $\Lambda_{\overline{\rm MS}}^{(5)}$ are expressed
in terms of $\alsm$ in table~1. $\alsm$ allows a more direct comparison
of the different results, because $M_Z \gg m_{b}, m_{\gl}$.
Besides the quoted statistical error $\delta \alpha_s^{stat}$,
there is
a theory error due to the assumption on the factorization scale~\cite{FACT},
$\delta \alsm^{theor}$, which has been estimated
to be $\approx 0.005$.
This error dominates the statistical error
determined by the analysis.
{}From direct measurements at LEP one obtains
$\alsm = 0.122 \pm 0.006$~[1b].
Comparing the results on $\alsm$ given in table 1 with the
LEP result, we derive
the confidence levels of 8~\%, 70~\%, and 5~\%
for the hypotheses $N_{\gl} = 0,1,~{\rm and} \ 2$,
respectively, for the agreement of both values.
Here
$\delta \alpha_s^{theor} = 0.005$ was added
to the corresponding statistical errors in quadrature in
the definition of $\chi^2$.

The resulting up quark, gluon and gluino
distributions at $Q^2 = 10 {\GeV}^2$ and $100 {\GeV}^2$
of QCD, SUSY QCD with
a massless, 3 and 5 GeV gluino
are shown in Fig. 2a--f.  The distributions
$u_{SUSY}(x)$ and $g_{SUSY}(x)$ are softer at $Q^2 \le 10 \ GeV^2$,
the faster running coupling of QCD compensating for the initial
depletion of larger $x$ distributions in the SUSY case.
At small $x$, $x < .01$, the sea and gluon
distributions in the SUSY case are enhanced by the
splitting of larger $x$ gluinos.

The fit results are not very sensitive to the parameters of
the gluino distribution and allow large distributions at small
$x$ where the data is not yet very precise however.  The gluino
distribution is forced by the evolution equations
to behave like a sea quark, the existing DIS data forces
it to be quite soft.
The magnitude
of the gluino distribution is strongly dependent on
$m_{\gl}$ over a large range of $Q^2$, the dependence
decreases logarithmically with increasing scale.

If Majorana gluinos exist in the mass range $m_{\gl} < 5$ GeV,
they can be searched for in single and pair production
in deep inelastic $ep$--scattering and photoproduction at
HERA.  The process of single gluino production will moreover
allow the derivation of direct constraints on the gluino distribution.

\vspace*{5mm}

\noindent{\bf{Acknowledgements}}\\
\noindent The authors would like to thank the members of
the $CTEQ$ collaboration, especially Wu-Ki Tung,
for sharing their expertise in QCD global analysis of
parton distributions.

\newpage

\newpage
\begin{table}[h]
\begin{center}
\begin{tabular}{|c|c||r|c|c|rcr|r|}
\hline \hline
\multicolumn{1}{|c|}{$N_{\gli}$}                               &
\multicolumn{1}{c||}{$m_{\gli} / \GeV$}                         &
\multicolumn{1}{c|}{$\chi^2$}                                  &
\multicolumn{1}{c|}{$N_{data \ points}$}            &
\multicolumn{1}{c|}{$\chi^2$/NDF}                                    &
\multicolumn{3}{c|}{$\Lambda_{\overline{\rm MS}}^{(5)} \pm
\delta \Lambda^{stat}$}                                               &
\multicolumn{1}{c|}{$\alsm \pm \delta \alpha_s^{stat}$}
\\
\hline \hline
0 &     & 783 & 986
& 0.808 & 118.0 &$\pm$& 13 MeV &  0.108 $\pm$ 0.002  \\
\hline
1 & 0   & 772 & 986
& 0.799 &   7.7 &$\pm$& 0.65 MeV & 0.124 $\pm$ 0.001  \\
1 & 3   & 772 & 986
& 0.797 &   7.7 &$\pm$& 0.61 MeV & 0.124 $\pm$ 0.001   \\
1 & 5   & 544 & 589 &
0.951 &  127.0&$\pm$& 4.7  MeV & 0.125 $\pm$ 0.001 \\
\hline
2 & 0   & 770 & 986
& 0.797 &  3.7  &$\pm$& 2.5 keV  & 0.145 $\pm$ 0.003 \\
2 & 3   & 786 & 986
& 0.811 &  7.7  &$\pm$& 6.5 keV  & 0.153 $\pm$ 0.009   \\
2 & 5   & 538 & 589
& 0.941 & 123.0 &$\pm$& 12 MeV   & 0.145 $\pm$ 0.009 \\
\hline \hline
\end{tabular}
\end{center}
\caption[xxx]{Results of the QCD analysis for $\Lambda$ and $\alsm$}
\end{table}

\begin{table}[h]
\begin{center}
\begin{tabular}{|c||r|r|r|}
\hline \hline
\multicolumn{1}{|c||}{          }                               &
\multicolumn{1}{c|}{QCD}                                       &
\multicolumn{2}{c|}{Majorana gluino}  \\ \cline{3-4}
\multicolumn{1}{|c||}{Parameters}                               &
\multicolumn{1}{c|}{   }                                       &
\multicolumn{1}{c|}{$m_{\gli} = 0$} &
\multicolumn{1}{c|}{$m_{\gli} = 3 \GeV$} \\
\hline \hline
$\gamma_u$      & 2.120    & 2.061     & 2.084     \\
$\gamma_d$      & --1.167    & --1.344     & --1.292     \\
\hline
$A_G$           & 0.4742    &  0.4716    &  0.4709    \\
$\alpha_G$      & --0.3467  &  --0.3470    & --0.3519      \\
$\beta_G$       & 6.708 & 6.647 &  6.636      \\
$\gamma_G$      & 2.975    & 2.922     & 2.962     \\
\hline
$\alpha_-$      & 1.407    & 1.326     & 1.351     \\
$\delta_-$      & 26.42    &  19.71    & 21.72     \\
$\gamma_-$      & --22.90    &  --15.13 & --17.64   \\
\hline
$A_{\tilde{g}}$ & $\bullet$    & 0.2137      & $\bullet$\\
$\alpha_{\tilde{g}}$
                & $\bullet$  & 0.0001     & $\bullet$    \\
$\beta_{\tilde{g}}$
                & $\bullet$    & 11.34     & $\bullet$    \\
\hline
\multicolumn{4}{|c|}{relative normalizations} \\
\hline
SLAC                      & 1.002    & 0.998     & 0.999   \\
NMC $\sqrt{s} = 90 \GeV$  & 1.004    & 1.003     & 1.003     \\
NMC $\sqrt{s} = 280 \GeV$ & 1.016    & 1.010     & 1.013     \\
BCDMS                     & 0.984    & 0.978     & 0.981     \\
CCFR                      & 0.971    & 0.966     & 0.967     \\
ZEUS                      & 1.010    & 1.002     & 1.004     \\
H1                        & 0.961    & 0.952     & 0.956     \\
\hline \hline
\end{tabular}
\end{center}
\caption[xxx]{Results of the QCD analysis for the parameters
of the input distributions at $Q^2_0~=~2.56~\GeV^2$}
\end{table}
\newpage

\epsfig{file=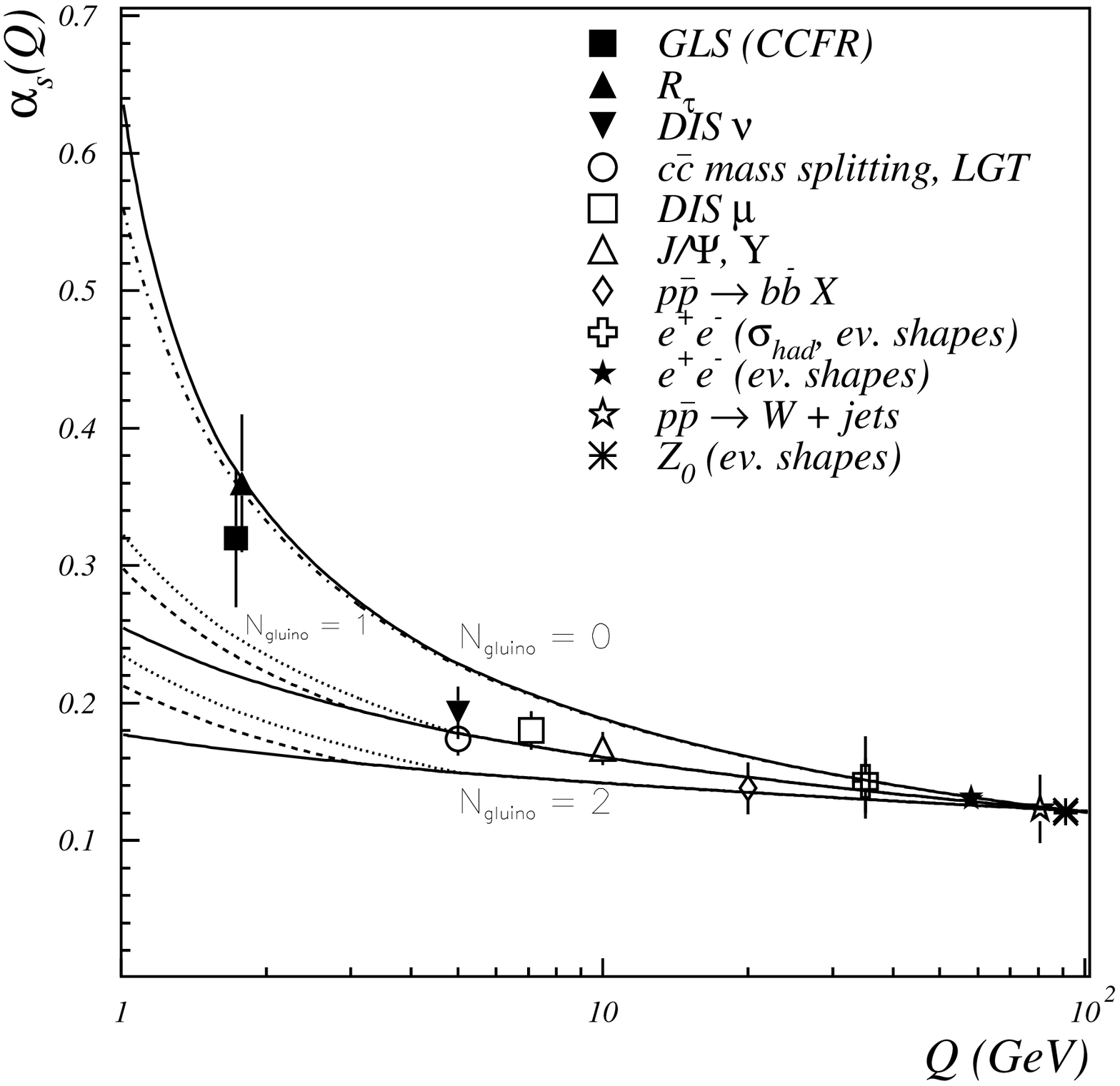,height=19cm,width=18cm}

\noindent
\small
Figure 1:~Comparison  of different theoretical predictions
for  $\alsq$ with  experimental
results of $\alpha_s$~\cite{REV}. The full curves denote the NLO solution
of eq.~(\ref{eq1}) for $N_{\gli} = 0,1,2$ with $m_{\gli} = 0$
taking
$\alsn = \alsm = 0.122$. The dash--dotted  line denotes the NNLO solution
in the case of QCD. The dashed and dotted lines describe the cases
$m_{\gli} = 3~{\rm and}~5 \GeV$, respectively.
\normalsize
\newpage
\noindent{\huge{a)}

\vspace*{-.4in}

\epsfig{file=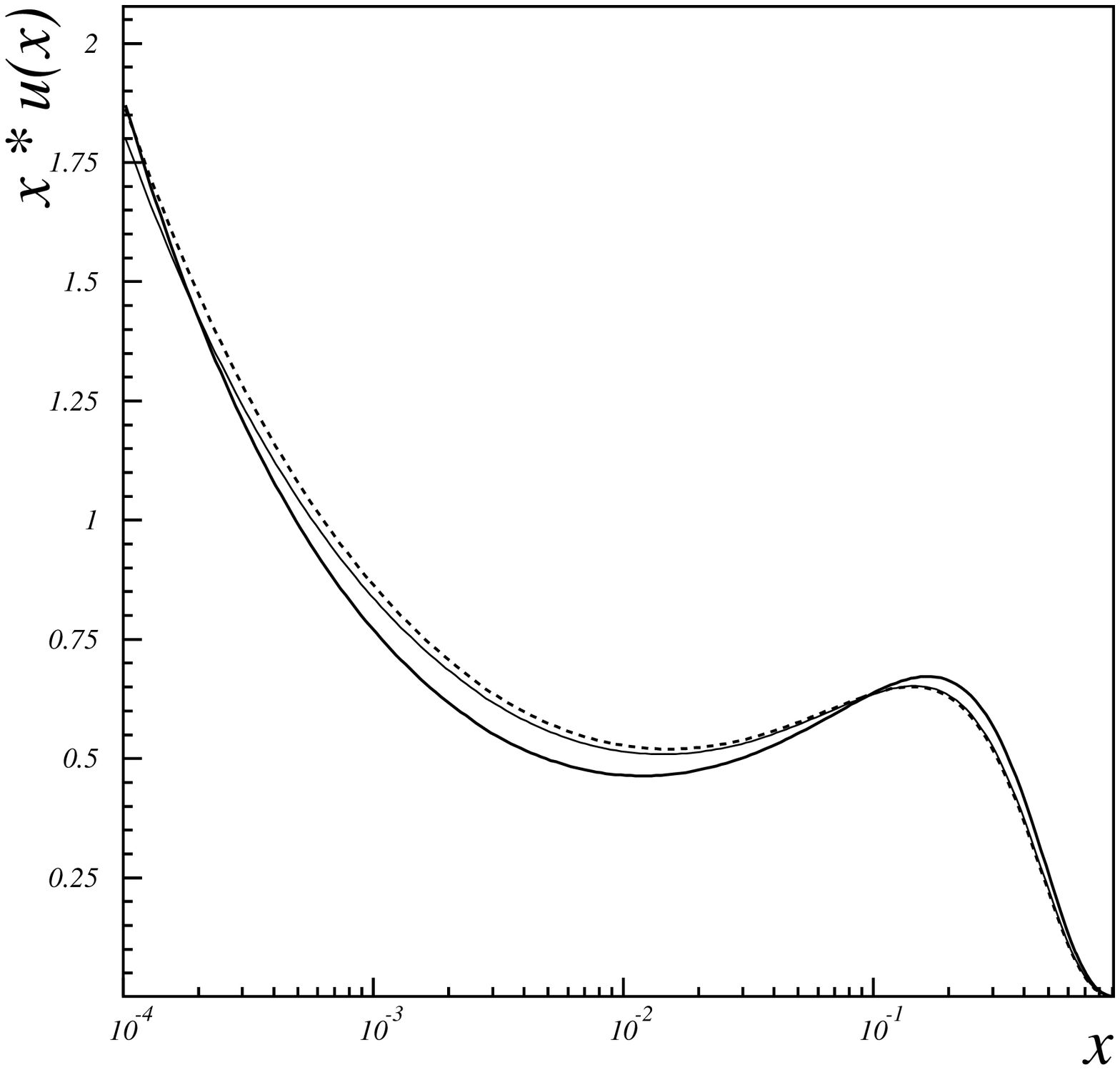,height=19cm,width=18cm}

\noindent
\small
Figure 2:~Fit results for the parton distributions. a)~$xu(x,Q^2)$,
b)~$xG(x,Q^2)$, c)~$\gli(x,Q^2)$ for $Q^2 = 10 \GeV^2$. The thick
solid, thin solid and dashed lines denote
QCD, SUSY QCD with
a massless Majorana gluino and SUSY QCD with
a Majorana gluino of $m_{\gl} = 3$ GeV respectively.
d)--f) are same as a)--c) with $Q^2 = 100 \GeV^2$.  The
dotted line denotes SUSY QCD with a Majorana gluino of
$m_{\gl} = 5$ GeV.
\normalsize

\newpage
\noindent{\huge{b)}

\vspace*{-.4in}

\epsfig{file=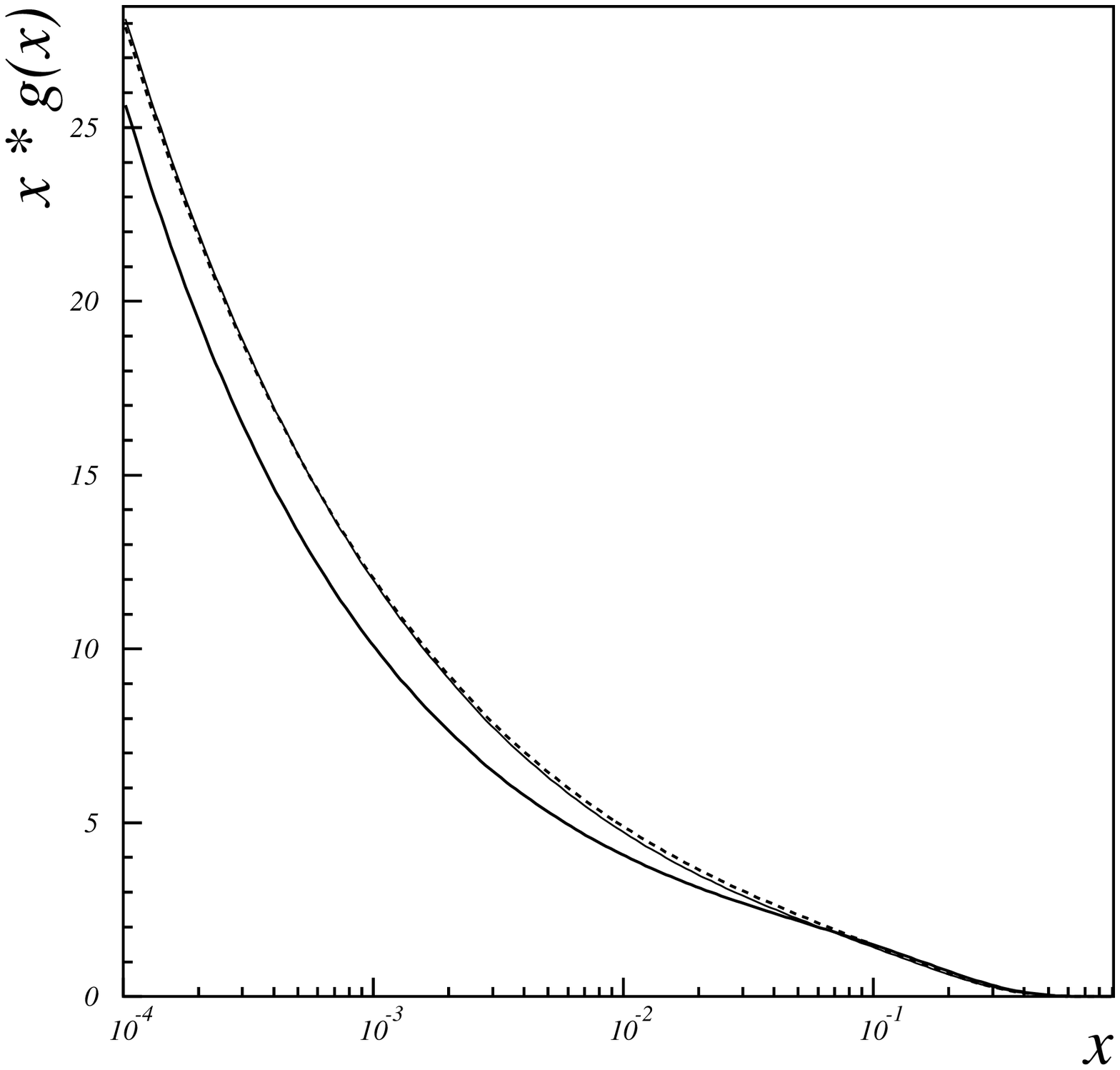,height=19cm,width=18cm}

\newpage
\noindent{\huge{c)}

\vspace*{-.4in}

\epsfig{file=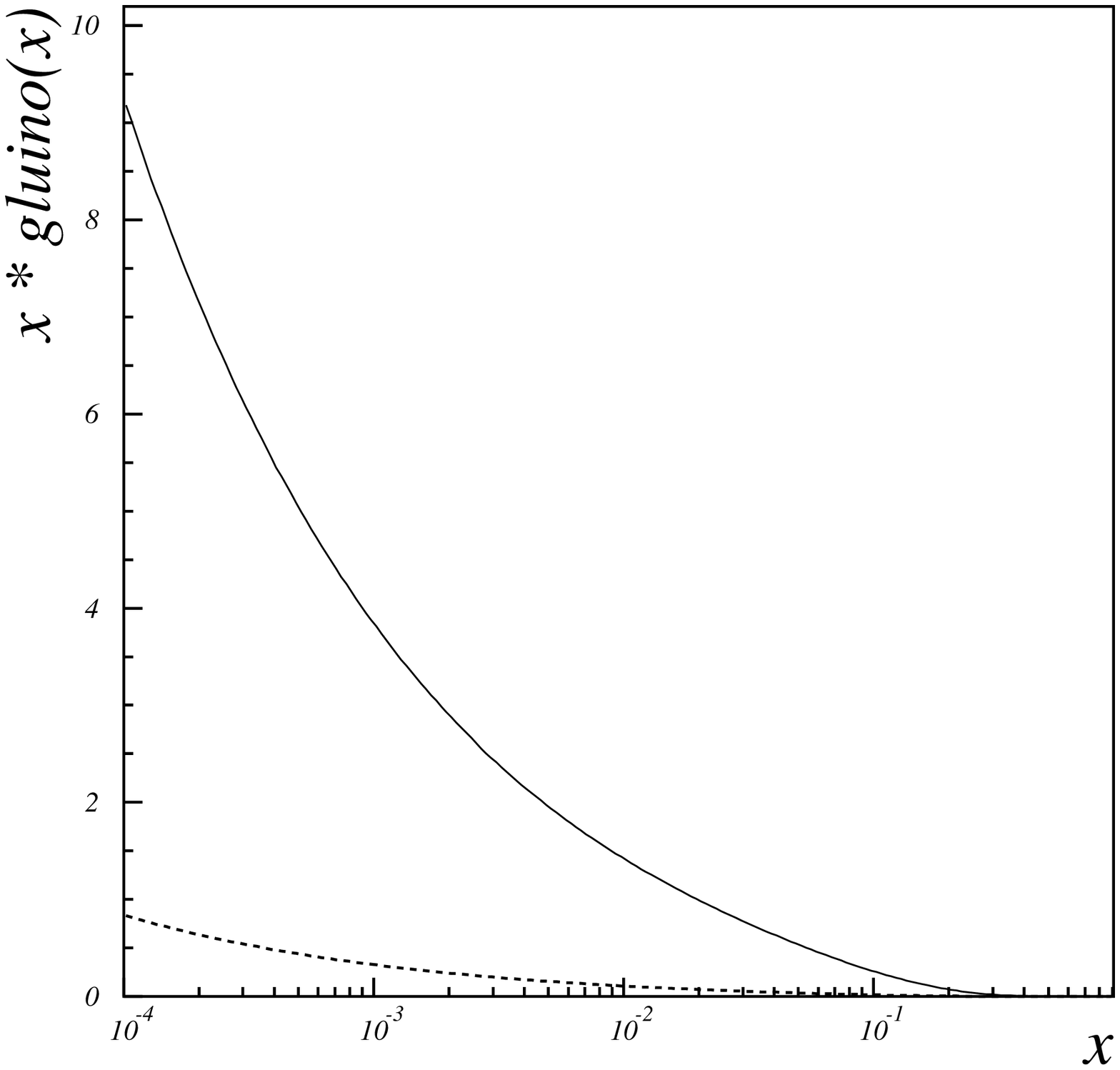,height=19cm,width=18cm}

\newpage
\noindent{\huge{d)}

\vspace*{-.4in}

\epsfig{file=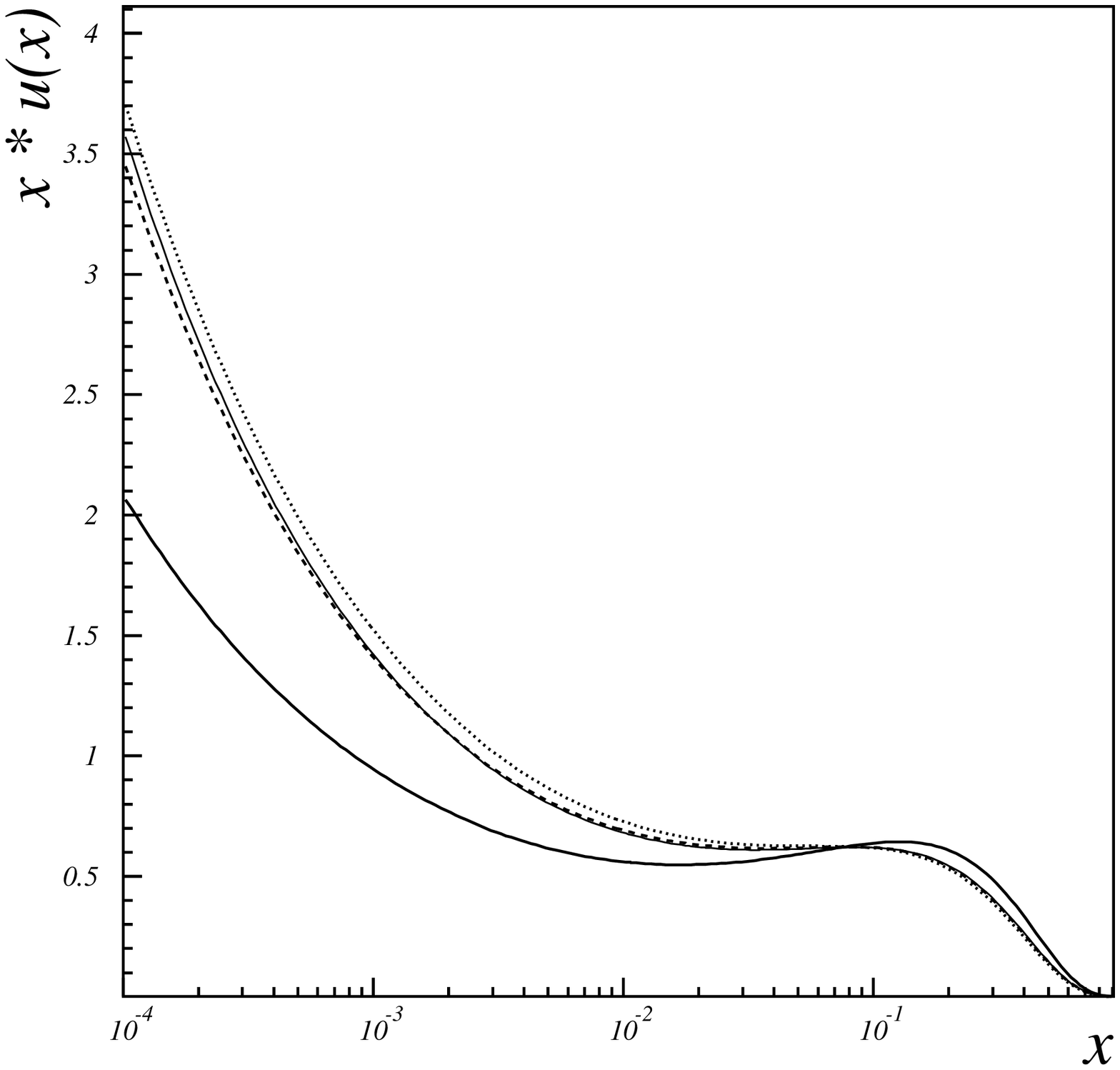,height=19cm,width=18cm}

\newpage
\noindent{\huge{e)}

\vspace*{-.4in}

\epsfig{file=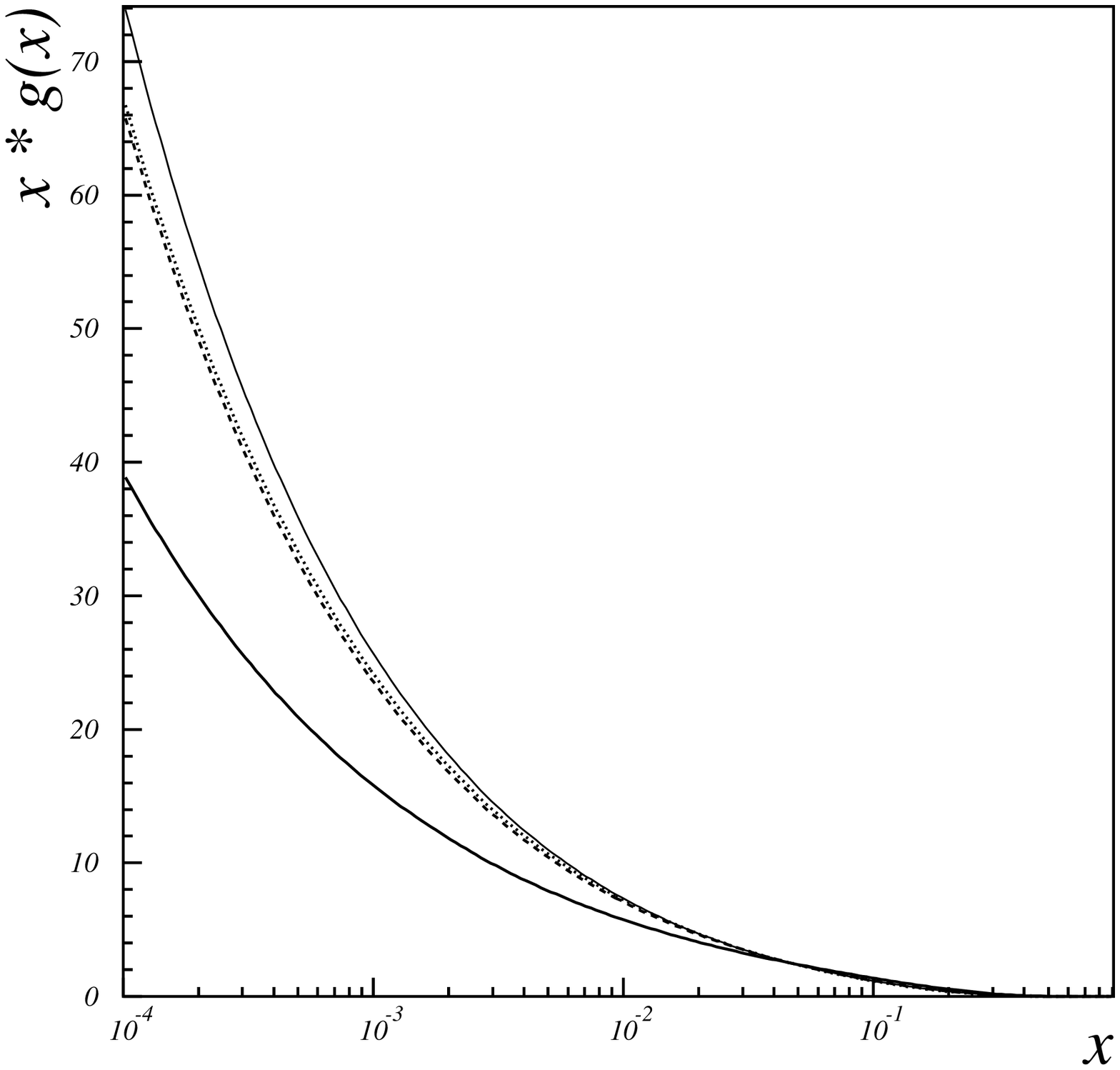,height=19cm,width=18cm}

\newpage
\noindent{\huge{f)}

\vspace*{-.4in}

\epsfig{file=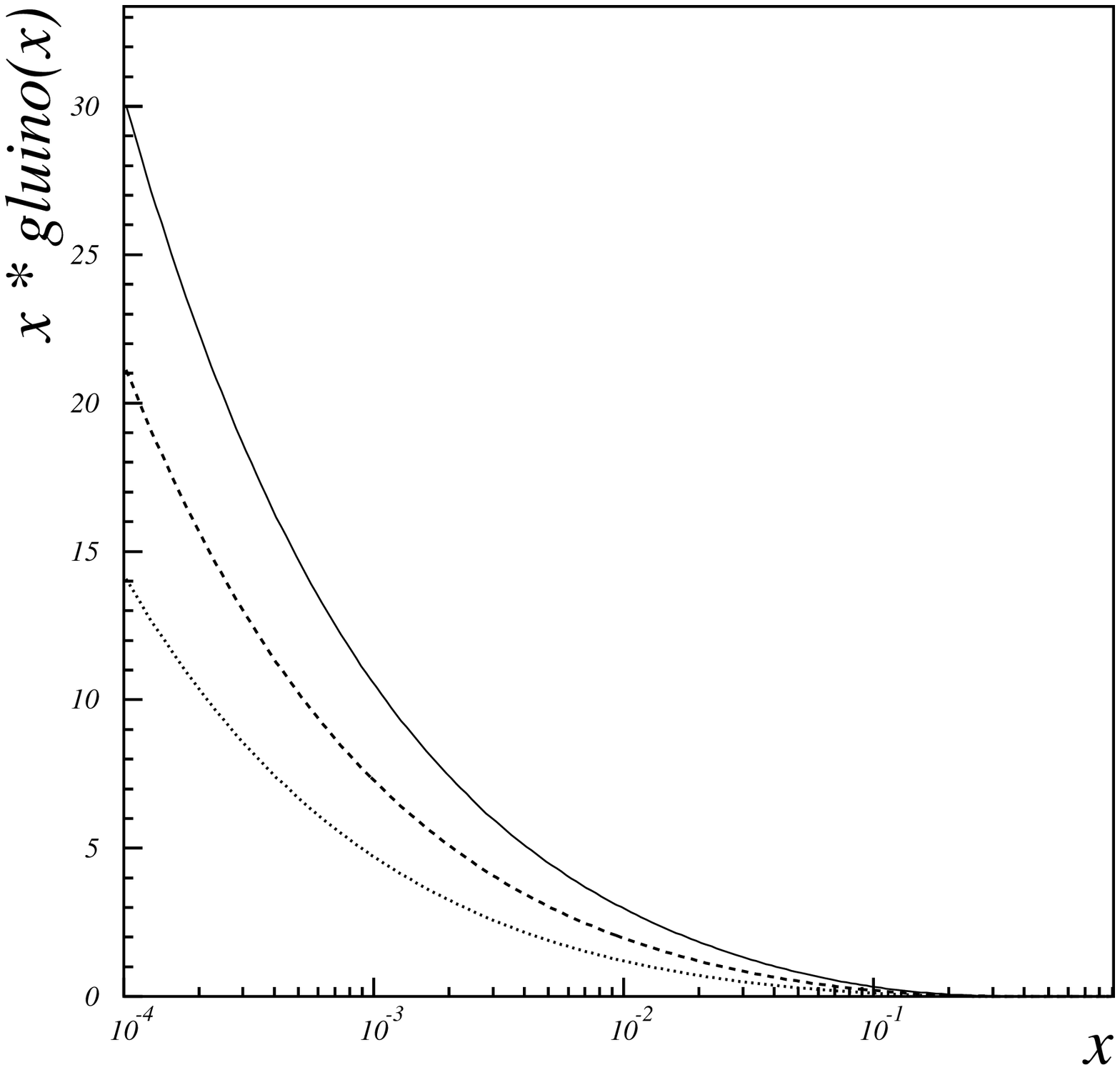,height=19cm,width=18cm}

\end{document}